\documentclass[reprint,amsmath,amssymb,showkeys]{revtex4-2}
\usepackage{CJK}
\usepackage{graphicx}
\usepackage{dcolumn}
\usepackage{bm}
\usepackage{float}
\usepackage[dvipsnames]{xcolor}
\usepackage{enumerate}

\newcommand{\kB}{k_\mathrm{B}}

\usepackage{xr}
\makeatletter

\begin{document}

\begin{CJK*}{}{} 
\title{Study of the uniform electron gas through parametrized partition functions}
\author{Tommaso Morresi$^{1,2\ast}$, Giovanni Garberoglio$^{2}$, Hongwei Xiong$^{1}$, Yunuo Xiong$^{1\ast}$\\~\\}
\affiliation{$^{1}$ Center for Fundamental Physics and School of Mathematics and Physics, Hubei Polytechnic University, Huangshi 435003,
People's Republic of China}
\affiliation{$^{2}$ European Centre for Theoretical Studies in Nuclear Physics and Related Areas (ECT*), Fondazione Bruno Kessler, Italy}

\begin{abstract}
We investigate the energy per particle, static structure factor, and momentum distribution of the uniform electron gas for different conditions defined by the dimensionless temperature $\Theta = 0.25 - 1.0$ and average interparticle distance $r_s = 0.5 - 80.0$ using path-integral Monte Carlo (PIMC) simulations. For small $r_\text{s}$ ($r_\text{s}\leq10$) where the sign problem is particularly challenging, we employ a recent approach based on an analytic continuation of the partition function using a real parameter $\xi$, which allows a generalization from bosons ($\xi=1$) to fermions ($\xi=-1$). We show that the results are in good agreement with other state-of-the-art methods while requiring low computational resources. For large $r_\text{s}$ ($r_\text{s}=80$), we use direct PIMC exploiting the good behaviour of the thermodynamic properties for negative $\xi$. In this framework we demonstrate that, for large $r_s$, the small negative region of $\xi$ can be utilized to extract information about the true fermionic limit, where $\xi = -1$. 
\end{abstract}

\maketitle
\end{CJK*}

\section{Introduction}
The uniform electron gas (UEG) is a fundamental model in condensed matter and plasma physics that describes a system of interacting electrons in a uniform neutralizing background. This model provides crucial insights into the behaviour of electrons in metals, semiconductors, and other many-body systems, serving as a cornerstone for understanding electronic correlations and exchange effects~\cite{Giuliani_Vignale_2005} and is therefore fundamental to atomic, molecular, and chemical physics, as well as materials science. 

The primary challenge associated with UEG is the Fermion sign problem (FSP), which arises from the antisymmetric nature of the electronic wavefunction when particle indices are exchanged~\cite{loh_1990,troyer_2005,dornheim_2_2018,dornheim_2019,bonitz2024toward}. For instance, in the context of PIMC simulations, the FSP causes an exponential increase in computational time as the number of particles increases and the temperature decreases. Consequently, it is essential to develop reliable and improvable approximations to mitigate the effects of the FSP. For the purpose of developing and testing computational methods, the UEG has been widely used as a benchmark. In density functional theory, the UEG underlies the local density approximation, a key building block of more sophisticated exchange-correlation functionals. Various forms of quantum Monte Carlo have emerged for both ground-state and finite-temperature simulations~\cite{ceperley1980ground,zhang2003quantum,booth2009fermion,brown_2013,blunt_2014,schoof_2015,groth_2016,dornheim_1_2016,yilmaz_2020,hirshberg_2020,dornheim2020attenuating,lee_2021,xiong_2022,prokof2007bold,hou2022exchange}. Two methods are particularly relevant to this work and will serve as a baseline to test our results: i) the configuration path integral Monte Carlo (CPIMC)~\cite{schoof_2011,schoof_2014} which provides exact results in the high-density region ($r_s \lesssim 5$, where $r_s=d/a_0$, $d$ is the average interparticle distance and $a_0$ the Bohr radius) of the UEG; and ii) the permutation blocking path integral Monte Carlo (PB-PIMC)~\cite{dornheim_1_2015,dornheim_2_2015}, which is widely considered the leading method for lower densities ($r_s \gtrsim 5$). The combination of these two methods enables accurate determination of the energy of the UEG over a wide range of densities for temperatures above half the Fermi energy.

In particular, recently the UEG has been widely studied in the warm dense matter (WDM) region~\cite{dornheim_2_2018,dornheim_1_2024,vorberger2025roadmap}. This region is characterized by a combination of high densities, corresponding to $r_s \sim 1$, and high temperatures, corresponding to $\Theta \sim 1$ where $\Theta = \frac{k_\text{B}T}{E_\text{F}}$, with $k_\text{B}$ being the Boltzmann constant, $T$ the temperature and $E_\text{F}$ the Fermi energy~\cite{graziani_2014}.
In fact, WDM plays a critical role in processes prior to inertial confinement fusion that occur in the target material when subjected to extreme conditions, such as those created by high-power lasers~\cite{hu_2011,gomez_2014,vorberger2025roadmap}. Notably, recent advances in understanding UEG \cite{bonitz2024toward,dornheim_1_2024,dornheim_1_2023,vorberger2025roadmap} in the WDM regime have been facilitated by the newly proposed framework of fictitious identical particles~\cite{xiong_2022}. This strategy \cite{xiong_2022,dornheim_1_2023,dornheim_1_2024,dornheim2024ab, dornheim2024ab1, dornheim2024unraveling,xiong_2023,morresi_2025} is based on the observation that the quantum statistical partition functions for bosonic and fermionic systems differ only by a single real parameter $\xi$: for bosons, $\xi=1$; whereas for fermions, $\xi=-1$. Since the partition function turns out to be continuous with respect to $\xi$, it is possible to devise an extrapolation method for the fermionic limit, based on numerical results from the $\xi \geq 0$ range. In the bosonic sector ($\xi > 0$), PIMC simulations are not affected by the sign problem and can be performed exactly. However, the constant-temperature $\xi$-extrapolation method~\cite{xiong_2022,dornheim_1_2023} has been shown to break down in the regime of strong quantum degeneracy~\cite{dornheim_1_2023}.

In this paper, we examine both the constant-temperature $\xi$-extrapolation method~\cite{xiong_2022,dornheim_1_2023,dornheim_1_2024,dornheim2024ab, dornheim2024ab1, dornheim2024unraveling} and the more general constant-energy extrapolation \cite{xiong_2023,morresi_2025} to study the UEG under average and strong quantum degeneracies across different densities. In particular, we show that the constant-energy extrapolation and the developed tailored extrapolation strategy in Ref.~\cite{morresi_2025} for normal liquid $^3$He allow to recover exact results for many different thermodynamic conditions. 
By using an example of UEG at low density, we also emphasize the importance of the small negative $\xi$ region, where the system always obeys an analytical behaviour. Notably, the sign problem is generally less pronounced in this region, and a few points with small negative $\xi$ values may suffice for effectively extrapolating physical information to the fermionic limit at $\xi = -1$. This characteristic can be extremely important when dealing with more complex scenarios.

To further support our findings, we also illustrate the distinct behaviors of the UEG at varying densities and temperatures by employing two independent particle models. One model approximates the single-particle dynamics within the Hartree--Fock (HF) approach, while the other addresses collective excitations, such as plasmon modes. At high density, it is well established that electrons in the UEG behave nearly like a non-interacting gas, leading to a smooth analytic relationship of energy as a function of $\xi$. However, as the density decreases, quantum many-body correlation effects become increasingly important, leading to non-analytic behavior in the energy as a function of $\xi$ in the region where $\xi>0$. In this work, we will demonstrate how the strategy proposed in Ref. \cite{morresi_2025} can be used to overcome this difficulty.\\
\indent The paper is organized as follows. In Sec. \ref{TF}, we introduce the parametrized partition function and independent particle models for the fictitious identical particles. In Sec. \ref{CD}, we present the technical details of the PIMC simulations used to compute thermodynamic and structural properties of the fictitious identical particles, and describe the constant-energy extrapolation method employed in this work to infer the fermionic energy. In Sec. \ref{Results}, we present a series of simulation results for the uniform electron gas at different densities. A brief summary and discussion are given in Sec. \ref{DC}.

\section{Theoretical framework}
\label{TF}
\subsection{Parametrized partition function for fictitious identical particles}
All the calculations for the UEG in the present work are performed using direct PIMC + worm algorithm calculations~\cite{dornheim_2019,ceperley_1995,boninsegni_2006}. The key details of the algorithm can be found in previous works~\cite{spada_2022,morresi_he4,morresi_2025}. 
Here, we only briefly mention that the parametrized partition function in the canonical ensemble for a system comprising $ N_{\uparrow} $ spin-up and $N_\downarrow$ spin-down electrons, such that $N=N_\uparrow+N_\downarrow$, in a cubic box subject to periodic boundary conditions, is expressed as follows:
\begin{equation}\label{eq:zxi}
\small
  Z(T,\xi)=\frac{1}{N_{\uparrow}!N_{\downarrow}!}\sum_{\mathcal{P_{\uparrow}}\mathcal{P_{\downarrow}}} \xi^{N_\mathcal{P}} \sum_{\mathbf{W}} \int_{\Omega} \text{d} \mathbf{X} \ \langle \mathbf{X}, 0 | \text{e}^{- \beta \hat H} | \mathcal{P_{\uparrow}} \mathcal{P_{\downarrow}}  \mathbf{X}, \mathbf{W} \rangle,
\end{equation}
where $\beta = (\kB T)^{-1}$, $\hat H$ is the Hamiltonian we will introduce below, $\xi$ is a real parameter, $\Omega$ is the volume of the cubic box, $N_{\mathcal{P}}$ denotes the minimum number of times for which pairs of indices have to be interchanged in the current permutation $\mathcal{P}=\mathcal{P}_\uparrow \mathcal{P}_\downarrow $ to recover the identity permutation and $ | \mathbf{X}, \mathbf{W} \rangle \equiv | \mathbf{R} \rangle$ is a general position state where $\mathbf{X}=\left(\mathbf{r}_1, s_1; \mathbf{r}_2, s_2; \dots,\mathbf{r}_N, s_N\right)$ is a compact notation for all the particle coordinates, that is positions $\mathbf{x}_i$ indicates the coordinates inside the simulation box as well as spins $s_i$ while $\mathbf{W}=(\mathbf{w}_1,\mathbf{w}_2,\dots,\mathbf{w}_N)$ is a vector of integers which can take any value in ($-\infty$,$\infty$) that takes into account the periodic images of the system. In the cases where $\xi=1$ and $\xi=-1$ one can recognize the exact fermionic and bosonic partition functions, respectively. It is worth noting that in the worm algorithm of PIMC with periodic boundary conditions employed in this work~\cite{morresi_he4,spada_2022}, incorporating the parametrized partition function for fictitious identical particles, as given by Eq.~(\ref{eq:zxi}), is straightforward compared to the case of bosons, and does not pose any particular difficulty.

The energy of the system described by Eq.~(\ref{eq:zxi}) can be derived using standard statistical mechanics and is given by:
\begin{align}
    E_{\xi}(T) = -\frac{1}{Z_{\xi}(T)}\frac{\partial Z_{\xi}(T)}{\partial \beta},  
\end{align}
where the subscript $\xi$ indicates that the variable $\xi$ is held constant. In order to reduce fluctuations, we employ the virial estimator (see Sec. S.1. in the Supplemental Material~\cite{supp}). To estimate the energy in the limit of fermionic systems, one can take advantage of the fact that both $Z_\xi(T)$ and $E_\xi(T)$ are continuous functions of $T$ and $\xi$. By examining their behavior in the region where $\xi \geq 0$, where calculations can be performed without encountering the sign problem, one can then extrapolate the energy value to $\xi = -1$. More details about the idea can be found in Refs.~\cite{xiong_2022, xiong_2023, dornheim_1_2023,dornheim_1_2024,dornheim2024ab, dornheim2024ab1, dornheim2024unraveling,morresi_2025}.

\subsection{UEG parameters}
The thermodynamic conditions of the UEG are fully characterized by the already introduced dimensionless parameters $r_s$ and $\Theta$, representing the average interparticle distance and temperature respectively, and by:
\begin{itemize}
    \item $\zeta$ , which denotes the polarization of the system;
    \item $N$, which is the total number of electrons considered.
\end{itemize}
From these parameters, we derive the densities of spin-up and spin-down electrons given by  $n^{\uparrow} = \frac{\zeta}{\frac{4}{3}\pi (a_0 r_\text{s})^3}$ and $ n^{\downarrow} = \frac{1 - \zeta}{\frac{4}{3}\pi (a_0 r_\text{s})^3}$, respectively, with the total density defined as  $n = n^{\uparrow} + n^{\downarrow}$. Additionally, the volume of the cubic box is given by $\Omega = N \frac{4}{3} \pi (a_0 r_s)^3$. The free gas Fermi wave vectors for spin-up and spin-down electrons are expressed as $k^{\uparrow \downarrow}_\text{F} = \left( 6 \pi^2 n^{\uparrow \downarrow} \right)^{1/3}$, and the Fermi energy is calculated using $E_\text{F}^{\uparrow\downarrow}=\frac{\hbar^2 {k_{\text{F}}^{\uparrow\downarrow}}^2}{2 m_\mathrm{e}}$, where $m_\mathrm{e}$ is the electron mass. For the sake of simplicity, in the following we will use atomic units unless otherwise stated. 

\subsection{Independent particle model}
\begin{figure}[!hbt]
\centering
\includegraphics[width=0.5\textwidth]{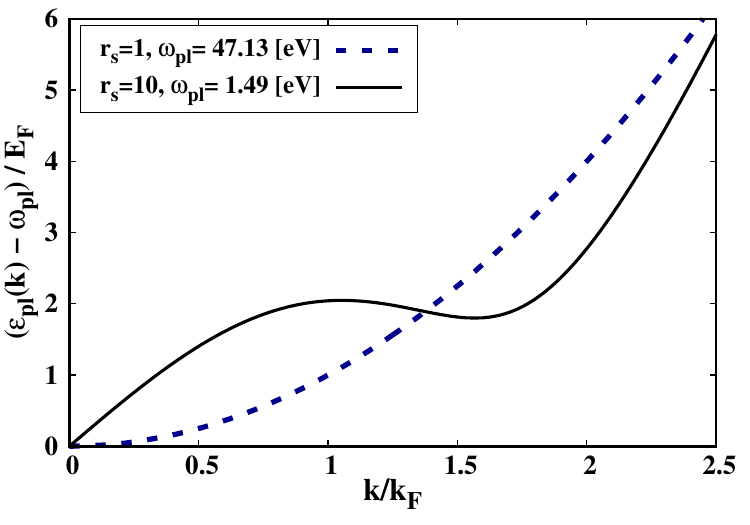}
\caption{Dispersion used for the plasmon in the independent particle model for $\zeta=1$ and for $r_s=1$ and $r_s=10$, respectively. The blue dashed line corresponds to the case $r_s=1$, while the continuous black line corresponds to the case $r_s=10$.
}\label{fig:dispersion}
\end{figure}

To gain a qualitative understanding of the energy behavior in the UEG, we numerically solve an independent particle model, as outlined in Ref.~\cite{morresi_2025}. This model is based on a single-particle dispersion relation that aims to qualitatively capture the physics of the system, resembling the solutions found in a mean-field theory. The chosen dispersion relation is connected to the shape of the dynamical structure factor (DSF) $S(\mathbf{k}, \omega)$. Specifically, the function $\omega(k)$ describing the dispersion relation is intended to describe the behavior of the peak in $S(\mathbf{k}, \omega)$. However, we emphasize that modeling the many-body properties of the DSF of the UEG using a single expression across the entire thermodynamic range of densities and temperatures is challenging. Indeed, the electronic $S(\mathbf{k},\omega)$ significantly alters its shape when transitioning from high-density to low-density regions~\cite{dornheim_1_2018,hamann_2020,dornheim_1_2022,chuna_2025,robles2025pylit,chuna2025secondroton}. For small momenta $k \ll k_\text{F}$, the system is predominantly influenced by collective plasmon excitations, which appear as a peak near the plasmon frequency $\omega_{pl}$. Conversely, for large momenta $ k \gg k_\text{F}$,  $S(\mathbf{k},\omega)$ approaches the single-particle limit and exhibits a parabolic dispersion relation. However, in the vicinity of  $k \sim k_\text{F} $, a roton feature emerges as the  $r_s$ value increases. To effectively capture this intricate behavior, we propose a dispersion relation  $\epsilon_{pl}(k)$ that is expressed as follows:
\begin{equation}\label{eq:disp_helium-like}
    \epsilon_{pl} (k) = \omega_{pl} +  \sqrt{\alpha_1 k \cdot \text{sin}\left(\alpha_2  k\right) + \left( \frac{ k^2}{2}\right)^2},
\end{equation}
where $\omega_{pl}=\sqrt{4 \pi n}$, $\alpha_2=9.45$ a.u. and $\alpha_1$ is such that $\sqrt{\alpha_1\alpha_2} \sim 3.8\cdot10^5 r_s^{1/3}$ a.u.. 
The parameters $\alpha_1$ and $\alpha_2$ are chosen to qualitatively reproduce the PIMC results for the DSF in the range of $r_s \in [1,10]$. The resulting dispersions in the limiting values of $r_s=1$ and $r_s=10$ are shown in Fig.~\ref{fig:dispersion} (dashed blue line and continuous black line, respectively). 

Conversely, the HF approximation provides a single-particle dispersion as a solution to the UEG problem. However, this solution exhibits a divergent derivative of the energy dispersion at $ k = k_\text{F}$~\cite{gellmann_1957}. To address this limitation, we turn to the hyper-Hartree--Fock (HHF) solution for the UEG~\cite{blair_2015}, which serves as a generalization of the HF approach. For ease of discussion, we will refer to the HHF solution simply as HF. Specifically, to model the HF single-particle energies, we employ the expression provided in Ref.~\cite{blair_2015}, which is outlined as follows:

\begin{equation}\label{eq:disp_hf}
    \epsilon_{HF}(k) = \frac{k^2}{2} - \frac{k_\text{F}}{\Lambda^{2/3}\pi} \left( 1+\frac{\Lambda^{2/3}k_\text{F}^2-k^2}{2\Lambda^{1/3}k_\text{F}k}\text{ln}\left| \frac{\Lambda^{1/3}k_\text{F}+k}{\Lambda^{1/3}k_\text{F}-k}\right|\right),
\end{equation}
where $\Lambda$ is an adimensional parameter that allows to recover the free particle dispersion ($\Lambda \to \infty$) and the pure Hartree-Fock solution ($\Lambda =1$). In this work, we set the parameter $\Lambda = \frac{3}{2}$; however,  the overall characteristics of the results presented herein are independent of this specific choice.

We considered the two single-particle dispersion models in Eqs.~(\ref{eq:disp_helium-like}) and (\ref{eq:disp_hf}) separately and solved the system independently. The main advantage of studying an independent particle model is that the relation between $E$, $\xi$ and $T$ can be computed numerically exactly. Furthermore, the independent particle model serves as a useful reference for guiding reliable $\xi$-extrapolation, as shown clearly in Ref. \cite{morresi_2025} for the $\xi$-extrapolation of normal liquid $^3$He.

In Ref.~\cite{Isakov93}, it is shown that for a generalized independent particle Hamiltonian where $H = \sum_\mathbf{k} \epsilon(\mathbf{k}) \hat{a}^\dagger_\mathbf{k}  \hat{a}_\mathbf{k}$ and where creation and annihilation operators are such that $ a_\mathbf{k} a^\dagger_{\mathbf{k}'} - \xi a^\dagger_{\mathbf{k}'} a_\mathbf{k} = \delta_{\mathbf{k}, {\mathbf{k}'}}$, occupation numbers have the same form of the usual fermion and bosonic occupations with $\xi$ in place of the $+1$ and $-1$ respectively,
\begin{equation}
    f(\epsilon; T, \mu) = \frac{1}{\exp\left(\frac{\epsilon-\mu}{T}\right) - \xi},
    \label{eq:occupation_numbers}
\end{equation}
where $\mu$ is the chemical potential.
In the grand-canonical ensemble the two equations relating the temperature $T$ and the chemical potential $\mu$ to the number of particles $N$ and the average energy $E$ are:
\begin{align}\label{eq:gc_N}
        N &= N_\mathrm{c} + \frac{\Omega}{(2\pi)^3} \int_0^{\infty} \frac{4 \pi k^2}{\text{exp}\left( \frac{\epsilon(k)-\mu}{T} \right) -\xi}\mathrm{d}k, \\
        \label{eq:gc_E}
        E &= \frac{\Omega}{(2\pi)^3} \int_0^{\infty} \frac{4 \pi k^2 \epsilon(k)}{\text{exp}\left( \frac{\epsilon(k)-\mu}{T} \right) -\xi}\mathrm{d}k,
\end{align}
where we assume spherical symmetry so that the dispersion relation depends only on $k = |\mathbf{k}|$, $\epsilon(k>0)>0$. 
The symbol $N_\mathrm{c}$ in Eq.~(\ref{eq:gc_N}) represents the non-zero macroscopic number of particles in the ground state, which is also the signature of Bose--Einstein condensation (BEC)~\cite{Pitaevskii16}. The presence of BEC in this system is related to the fact that for any $\xi > 0$ the chemical potential has an upper bound, otherwise the occupation numbers of Eq.~(\ref{eq:occupation_numbers}) become negative. This upper bound for the chemical potential is $\mu_\mathrm{c} = - T \log\xi + \epsilon(k=0)$.
Conversely, if one considers Eq.~(\ref{eq:gc_N}) as a function of $\xi$ for constant temperature, there is a critical value of the parameter $\xi$ after which $N_\mathrm{c} > 0$ that can be computed as
\begin{align}\label{eq:xicrit}   
\xi_{\mathrm{c}} &= \frac{\Omega}{N (2\pi)^3} \int_0^{\infty} \frac{4 \pi k^2}{\text{exp}\left( \frac{\epsilon(k)-\epsilon(k=0)}{T} \right) -1}\mathrm{d}k.
\end{align}
At this critical value, the energy as a function of $\xi$ exhibits non-analytic behavior due to the increasing occupation of $N_\mathrm{c}$. 

\section{Computational Details}
\label{CD}
In the UEG, the Hamiltonian operator $\hat H$ is written as \cite{dornheim_2_2018}
\begin{align}\label{eq:ham}
 \hat H=-\frac{1}{2}\sum^N_{i=1} \Delta_i + \hat V(\mathbf{R}),
\end{align} 
where $\hat V$ is the Coulomb interacting potential between electrons, and electrons with the static positive background that ensures charge neutrality. 
The imaginary time propagator in Eq.~(\ref{eq:zxi}) is then approximated using the primitive factorisation such that
\begin{equation}\label{eq:primitive_approx}
  \langle \mathbf{X}, 0 | \text{e}^{- \tau \hat H} | \mathcal{P} \mathbf{X}, \mathbf{W} \rangle \eqsim \langle \mathbf{X}, 0 | \text{e}^{- \tau \hat K} \text{e}^{- \tau \hat V}  | \mathcal{P} \mathbf{X}, \mathbf{W} \rangle.
\end{equation}
Here $\hat K = \hat{H}-\hat{V}$ is the kinetic operator.

The interacting potential $ V(\mathbf{R})$ in Eq.~(\ref{eq:ham}) is treated in two different ways, depending on the size of the edge $L=\Omega^{1/3}$ of the cubic cell. For $L<12.5 a_0$, we use the exact Ewald potential $W_\text{EW}$ for the UEG that reads as~\cite{dornheim_2_2018}:
\begin{equation}\label{eq:ew}
\begin{split}
    W_\text{EW}(\mathbf{r}_1,\mathbf{r}_2) = & \frac{1}{\pi\Omega} \sum_{\mathbf{G}\ne0} \left(\frac{e^{-\frac{\pi^2|\mathbf{G}|^2}{\kappa^2}+2\pi i\mathbf{G}\cdot(\mathbf{r}_1-\mathbf{r}_2)}}{|\mathbf{G}|^2} \right) - \frac{\pi}{\kappa^2 \Omega} + \\
&\sum_{\mathbf{m}}\frac{\text{erfc}\left(\kappa |\mathbf{r}_1-\mathbf{r}_2+\mathbf{m}L|\right)}{|\mathbf{r}_1-\mathbf{r}_2+\mathbf{m}L|},
\end{split}
\end{equation}
where 
$\mathbf{G}=\frac{\mathbf{j}}{L}$ and $\mathbf{j}$ is a vector of integers, $\kappa$ is the free parameter of the Ewald potential and $\mathbf{m}$ is another vector of integer numbers. We notice that the potential in Eq.~(\ref{eq:ew}) does not depend on the value of $\kappa$, which can be exploited for optimization.
In the calculation of the total energy, one should take into account the term $C_\text{EW}=\lim_{\mathbf{r}_1\to\mathbf{r}_2} \left( W_\text{EW}(\mathbf{r}_1,\mathbf{r}_2) - \frac{1}{|\mathbf{r}_1-\mathbf{r}_2|} \right)\eqsim\frac{M}{r_\text{s}(4 \pi N)^{1/3}}$,
where $M = -4.09209$, which is the Madelung constant for jellium, describing the self-interaction of the Ewald summation in periodic boundary conditions~\cite{groth_2016}. The Hamiltonian in this case is thus~\cite{schoof_2015}:
\begin{align}
 \hat H_\text{EW}=-\frac{1}{2}\sum^N_{i=1} \Delta_i+\frac{1}{2}\sum^N_{i<j}W_\text{EW}(\mathbf{r}_i,\mathbf{r}_j) + \frac{N}{2}C_\text{EW}.
\end{align}
Since Eq.~(\ref{eq:ew}) involves a summation over both real space and reciprocal space vectors---where the convergence in reciprocal space is slow---we precompute the Ewald potential on a cubic grid consisting of $81 \times 81 \times 81$ points in our simulations. Such points are then interpolated using cubic splines that make the calculations much more efficient. 

For $L\geq 12.5 a_0$, we used instead the spherically averaged potential of Yakub and Ronchi, $W_{\text{YR}}$ \cite{yakub_2003,demyanov_2024}, which allows much faster simulations while keeping a very good accuracy, that is~\cite{dornheim_yr}:
\begin{equation}
    W_\text{YR}(r_{12}) =  
       \begin{cases}
        \frac{1}{r_{12}}\left(1+\frac{r_{12}}{2r_c} \left[ \left(\frac{r_{12}}{r_c}\right)^2-3 \right] \right), & \text{if } r_{12} \leq r_c\\
        0,  & \text{if } r_{12} > r_c
    \end{cases}
\end{equation}
where $r_{12}=|\mathbf{r}_1 - \mathbf{r}_2|$ and $r_c=L\left(\frac{3}{4\pi}\right)^{1/3}$. In this case the constant term has a different form $C_\text{YR}=-\frac{3(N+5)}{10r_c}$ and the Hamiltonian becomes:
\begin{align}
\small
 \hat H_\text{YR}=-\frac{1}{2}\sum^N_{i=1} \Delta_i+\frac{1}{2}\sum^N_{i<j}\sum_{\mathbf{m}}W_\text{YR}(\mathbf{r}_i,\mathbf{r}_j + \mathbf{m}L) + \frac{N}{2}C_\text{YR}.
\end{align}
The energy differences of the two potentials for $L=12.5a_0$ cannot be resolved within the Monte Carlo error bars~\cite{dornheim_yr}.

For each value of the $\xi$-parameter, the estimators used to evaluate the different quantities presented below, such as energy-per-particle and momentum distribution, are reported in section S.1 of the Supplemental Material~\cite{supp}.

In particular, the fermionic energy-per-particle from PIMC simulations are computed using the fictitious identical particle framework~\cite{xiong_2022}. Within this framework, the fermionic energy can be obtained using two methods: constant-temperature and constant-energy extrapolations. In the first method~\cite{xiong_2022,dornheim_1_2023} the energy is obtained by fitting with a parabolic polynomial the energies for $\xi\geq 0$:
\begin{equation}
    E(T,\xi) = c_0 + c_1 \xi + c_2 \xi^2, 
    \label{eq:E_extrap}
\end{equation}
for each temperature. In the second method \cite{xiong_2023,morresi_2025}, one considers $\xi$ as a function of both $T$ and $E$ by inverting Eq.~(\ref{eq:E_extrap}). 
In particular the following relation holds:
\begin{equation}\label{eq:dcsi_dt}
    \frac{\partial \xi_{E}(T)}{\partial T} \Bigg|_{T=0} = 0,
\end{equation}
which means that, when expanding $\xi$ as a function of $T$, the linear term is missing. Therefore, for a fixed energy, the relation between $\xi$ and $T$ can be expressed as:
\begin{equation}\label{eq:csi_exp_xiong}
    \xi_E(T) = a_0(E) + a_2(E) T^2 + \sum_{i>2} a_i(E) T^i.
\end{equation}
In this framework, one computes numerically $E(\xi)$ for several values of $T$ and then samples the function $\xi_E(T)$. By fitting the parameters $a_i$ in Eq.~(\ref{eq:csi_exp_xiong}), one can then get the temperature $T_E$ where the fermionic system attains energy $E$ by solving 
$\xi_E(T)=-1$. 

Nevertheless, in the strong quantum degeneracy case, an improved strategy has been proposed in Ref.~\cite{morresi_2025} to consider the BEC transition in the bosonic sector. Indeed, when quantum effects are dominant, both $E_T(\xi)$ and $\xi_E(T)$ functions can have non-analytic behaviour in presence of BEC transition. Furthermore, finite size effects play also a role in the final shape of the functions by smoothening the curves.
Therefore, by noticing that the behavior of $\xi_E(T)$ close to $\xi = 0$ is effectively linear~\cite{morresi_2025}, a new fitting function for $\xi_E(T)$ has been proposed:
\begin{equation}\label{eq:csi_exp_mine_linear}
\begin{split}
    \tilde{\xi}_E(T) = & \left( a_0(E) +a_2(E) T \right) \theta(T-T_\mathrm{c}) + \\
    &\left(b_0(E) + b_1(E) T +b_2(E)T^2 \right) \bar{\theta}(T-T_\mathrm{c}), 
\end{split}
\end{equation}
where $\theta$ is the Heaviside step function, $\bar{\theta}(T) = 1-\theta(T)$ and $T_\mathrm{c}$ is estimated as the temperature after which the non-linear form of $\xi_E(T)$ is apparent. It is important to notice that the linear term in Eq.~(\ref{eq:csi_exp_mine_linear}) is not meant to describe the physics for $T \to 0$, where $\xi_E$ can take values well below $-1$, but only as a simple and effective extrapolation method to $\xi = -1$. We stress that, for normal liquid $^3\mathrm{He}$ \cite{morresi_2025}, the fitting function given by Eq.~(\ref{eq:csi_exp_mine_linear}) performs significantly better than Eq.~(\ref{eq:csi_exp_xiong}), primarily because the phase transition occurring in the bosonic regime is not accounted for in Eq.~(\ref{eq:csi_exp_xiong}).

\section{Results}
\label{Results}
\subsection{Independent particle model}
\begin{figure}[!hbt]
\centering
\includegraphics[width=0.5\textwidth]{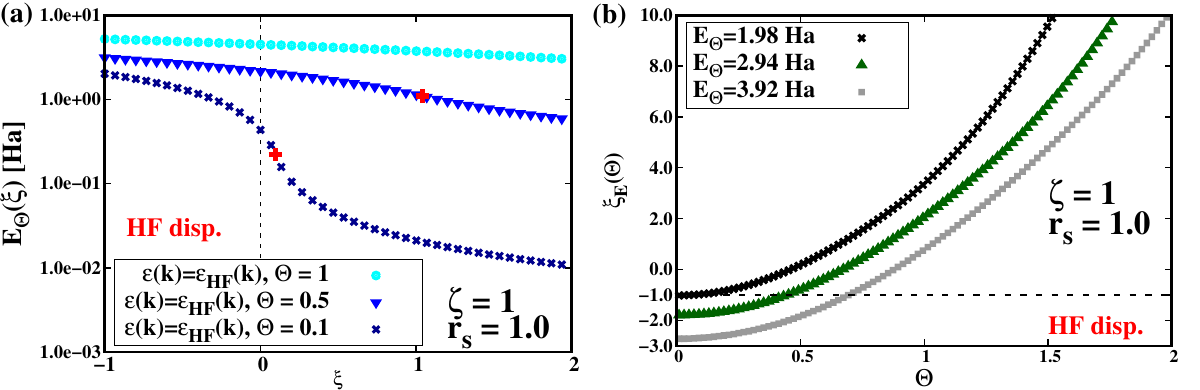}
\includegraphics[width=0.5\textwidth]{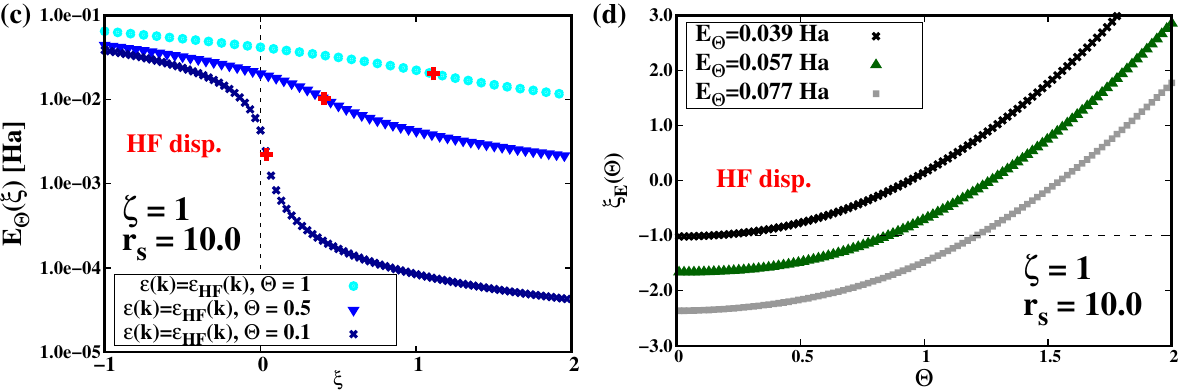}
\caption{Fully polarized independent particle model using the HF solution as dispersion in Eq.~(\ref{eq:disp_hf}). (a) Energy per particle as a function of $\xi$ for three different values of $\Theta$ in the case of $r_\text{s}=1.0$. The red points specify the energy at $\xi_\text{c}$ as computed from Eq.~(\ref{eq:xicrit}). (b) $\xi_\text{E}$-function for three fixed energies as function of $\Theta$ in the case of $r_\text{s}=1.0$. The intersection of these curves with the line $\xi=-1$ gives the temperature corresponding to the fixed energy. (c): same as (a) but for $r_\text{s}=10.0$. (d): same as (b) but for $r_\text{s}=10.0$.}\label{fig:ipm_hf}
\end{figure}

In Fig.~\ref{fig:ipm_hf}, we report the results of the independent particle model using the HF dispersion in Eq.~(\ref{eq:disp_hf}). In particular in Figs.~\ref{fig:ipm_hf}(a) and (c), we show three different energies-per-particle as a function of the $\xi$ parameter corresponding to three different values of $\Theta$ for $r_s=1$ and $r_s=10$, respectively. The red stars correspond to the critical points after which the system undergoes the BEC transition as computed in Eq.~(\ref{eq:xicrit}). We observe that at lower density ($r_s = 10$), the critical value $\xi_c$ is closer to $\xi = 0$ compared to the case at higher density ($r_s = 1$) at fixed temperature. Additionally, the drop to the ground state energy is also faster at lower density. As expected in both (a) and (c), the lower the temperature, the sharper the transition. In Figs.~\ref{fig:ipm_hf}(b) and (d) we show instead three constant-energy $\xi_E(\Theta)$ functions. In both $r_s=1.0$ and $r_s=10.0$ cases we varied the energies from $E_\text{F}$ (black points) to almost twice $E_\text{F}$ (grey points). We observe a smooth behavior of these functions, regardless of the energy and density of the system. Such a smooth behavior is reminiscent of the well-behaved case of a free particle (see the appendix of Ref.~\cite{morresi_2025}).

\begin{figure}[!hbt]
\centering
\includegraphics[width=0.5\textwidth]{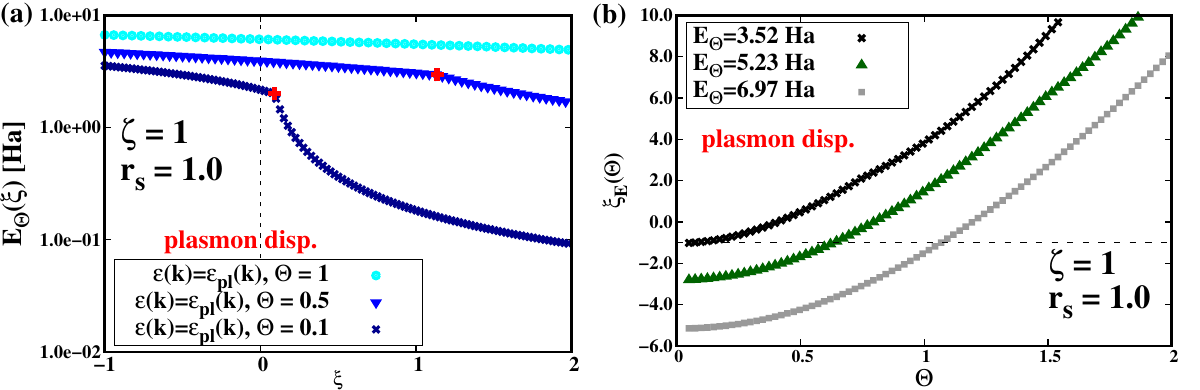}
\includegraphics[width=0.5\textwidth]{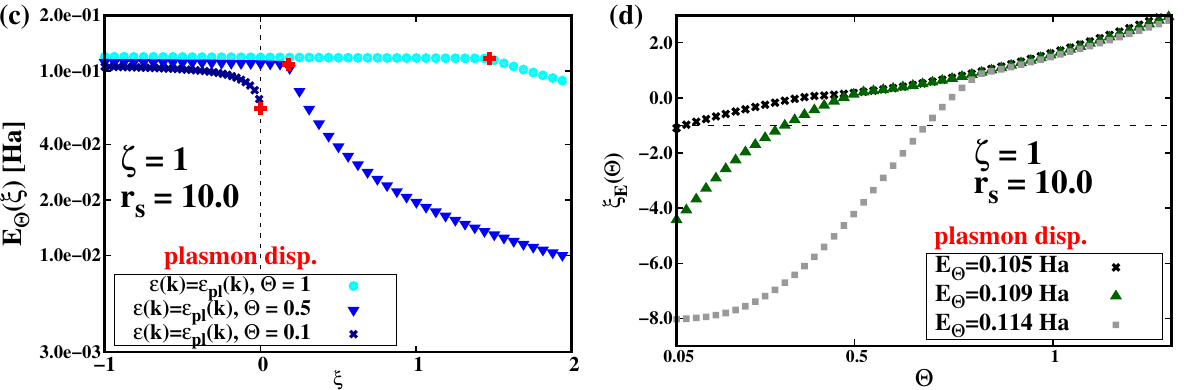}
\caption{Fully polarized independent particle model using the plasmon dispersion in Eq.~(\ref{eq:disp_helium-like}). (a) Energy per particle as a function of $\xi$ for three different values of $\Theta$ in the case of $r_\text{s}=1.0$. The red points specify the energy at $\xi_\text{c}$ as computed from Eq.~(\ref{eq:xicrit}). (b) $\xi_\text{E}$-function for three fixed energies as function of $\Theta$ in the case of $r_\text{s}=1.0$. The intersection of these curves with the line $\xi=-1$ gives the temperature corresponding to the fixed energy. (c): same as (a) but for $r_\text{s}=10.0$. (d): same as (b) but for $r_\text{s}=10.0$.}\label{fig:ipm_pl}
\end{figure}

In Fig.~\ref{fig:ipm_pl}, we present instead the results of the independent particle model using the plasmon dispersion described in Eq.~(\ref{eq:disp_helium-like}). In this case, the behavior of the energy per particle as a function of $\xi$, shown in Figs.~\ref{fig:ipm_pl}(a) and (c) for $r_s = 1.0$ and $r_s = 10.0$  respectively, is also characterized by sharp inflection points indicated by the red stars, corresponding to the solutions of Eq.~(\ref{eq:xicrit}). However, when using the plasmon dispersion, the change in curvature of the energy near the critical point is characterized by a singularity, resulting in a divergent first derivative $\frac{\partial E_{\Theta}}{\partial \xi}|_{\xi_c}=-\infty$. This singular behavior is more evident in Fig.~\ref{fig:ipm_pl}(c) and is reflected in the shape of the constant-energy $\xi_E(\Theta)$ functions shown in Figs.~\ref{fig:ipm_pl}(b) and (d). Specifically, while the behavior of these functions remains relatively smooth for $r_s = 1$ (see Fig.~\ref{fig:ipm_pl}(b)), in the case of $r_s = 10$ the BEC phase transition is reflected in the phase transition of the $\xi_E$. 
Based on this analysis, one can expect that the non-analytic behaviour of the $\xi_E$ in the UEG should become evident at low density ($r_s \geq 10$). Nevertheless, it is also expected that the inflection points in Figs.~\ref{fig:ipm_pl}(c) and (d) may be softened by finite-size effects~\cite{morresi_2025}. A more general picture on the behaviour of the energy as a function of $\xi$ and $\Theta$ is also shown in Fig.~\ref{fig_si:3d} of the Supplemental Material~\cite{supp} with the help of a heat map.

The analysis of the independent particle models indicates that, for the same $\Theta$, different fitting functions and strategies are required to infer the fermionic energy at different values of $r_s$. This observation is confirmed by our PIMC simulations presented below for $\Theta=0.5$, where quantum degeneracy effects are significant.

\subsection{PIMC results for UEG}
In the following subsections, we will present results for several UEG parameters, ranging from small to large values of $r_s$. Notably, we observe that for a fixed number of particles and temperature, the FSP become less pronounced as $r_s$ increases.
The selected set of parameters is designed to encompass the different regimes of the UEG, from behaviors resembling single-particle dynamics to those characteristic of strong correlations.

Additionally, we explore the region where FSP effects are manageable, specifically at $r_s = 80.0 $. In this case, we propose that the small negative range of $ \xi $ can be utilized to extract significant features of the true fermionic system.


\subsubsection{UEG for $r_\text{s}=0.5$, $\Theta=0.5$, $N=14$ and $\zeta=0$}\label{sec:rs_0.5}
As first case study, we consider the UEG for $r_\text{s}=0.5$, $\Theta=0.5$, $N=14$ and $\zeta=0$, as shown in Fig. 15 of Ref.~\cite{dornheim_1_2023}. Under such conditions, direct PIMC fails to reliably simulate the fermionic energy. 

\begin{figure}[!hbt]
\centering
\includegraphics[width=0.5\textwidth]{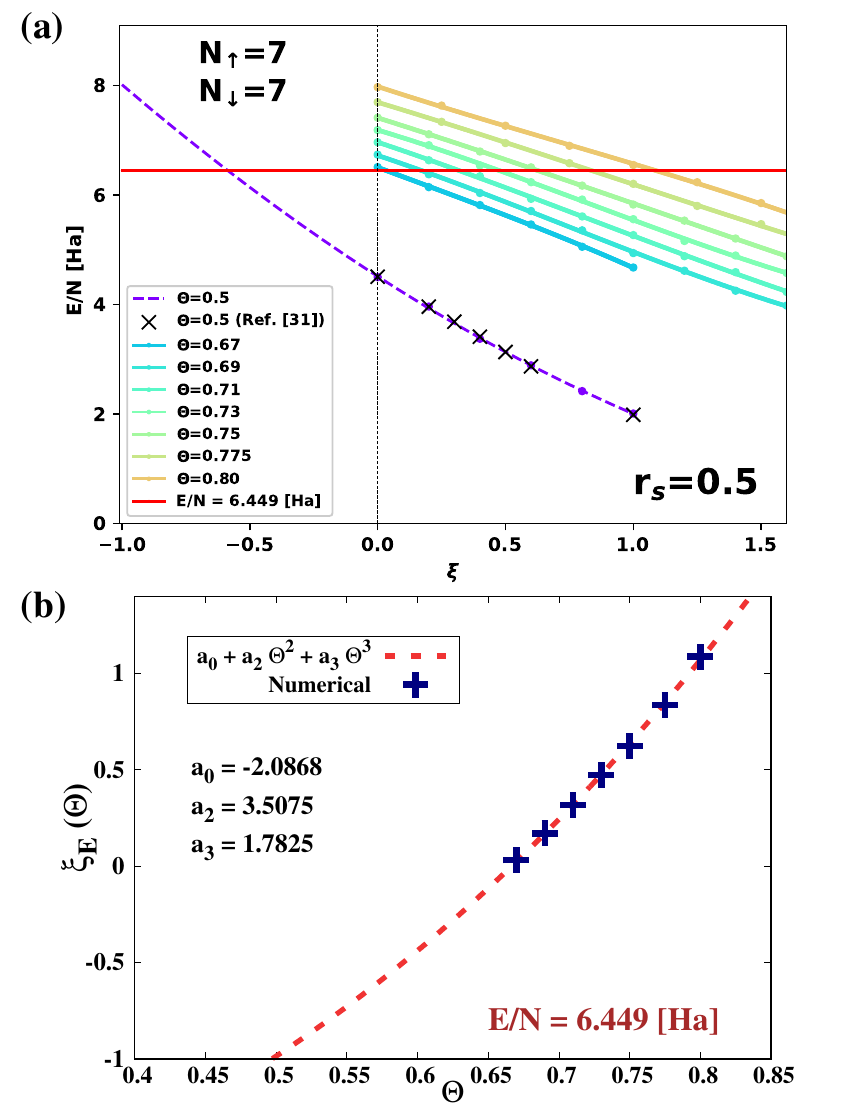}
\caption{(a) Energy per particle $r_\text{s}=0.5$, $N=14$ and $\zeta=0$ as a function of the $\xi$-parameter for several $\Theta$. Black crosses are taken from~\cite{dornheim_1_2023}, while the red line corresponds to energy per fermion ($\xi=-1$) using CPIMC at $\Theta=0.5$. Purple dashed line is the fit to the energy values for $\Theta=0.5$. (b) Constant energy extrapolation of $\xi_\text{E}(\Theta)$ obtained by cutting the $E_\text{T}(\xi)/N$ functions in (a) along the red line.}\label{fig:15_dorn}
\end{figure}

By building the $\xi_E(\Theta)$ function at $E/N=6.449$ Ha, corresponding to the CPIMC result for this set of parameters and shown by the red line in Fig.~\ref{fig:15_dorn}(a), one can obtain the blue points shown in Fig.~\ref{fig:15_dorn}(b). We fitted those points using Eq.~(\ref{eq:csi_exp_xiong}) up to the cubic term (red dashed line) and we found that the temperature corresponding to the chosen energy is $\Theta=0.5$, recovering thus the result of CPIMC. The uncertainty on the temperature is in this case less than $0.5\%$.

Instead, as it is shown in Ref.~\cite{dornheim_1_2023}, by comparing the exact result obtained using CPMIC with the one derived from the constant temperature $\xi$-extrapolation at $\Theta = 0.5$, an obvious discrepancy appears between the two values. This is also illustrated in Fig.~\ref{fig:15_dorn}(a), where the fit of the $\xi \geq 0$ points for $\Theta = 0.5$ (indicated by the purple dashed line) yields a value of $E(\xi = -1)/N$ that is noticeably different from the exact value of fermions represented by the red line.
This example clearly demonstrates the value of the constant-energy extrapolation procedure, which offers an improvement over the constant-temperature method suggested by Eq.~(\ref{eq:E_extrap}). This improvement arises because the constant-energy curves avoid the region where energy drops occur due to the BEC transition for $\xi > 0$ at $\Theta \leq 0.5$ (see Fig.~\ref{fig_si:3d} in the Supplemental Material~\cite{supp}.)

\subsubsection{UEG for $r_\text{s}=1.0$, $\Theta=0.5$, $N=33$ and $\zeta=1$}\label{sec:rs_1.0}
\begin{figure}[!hbt]
\centering
\includegraphics[width=0.5\textwidth]{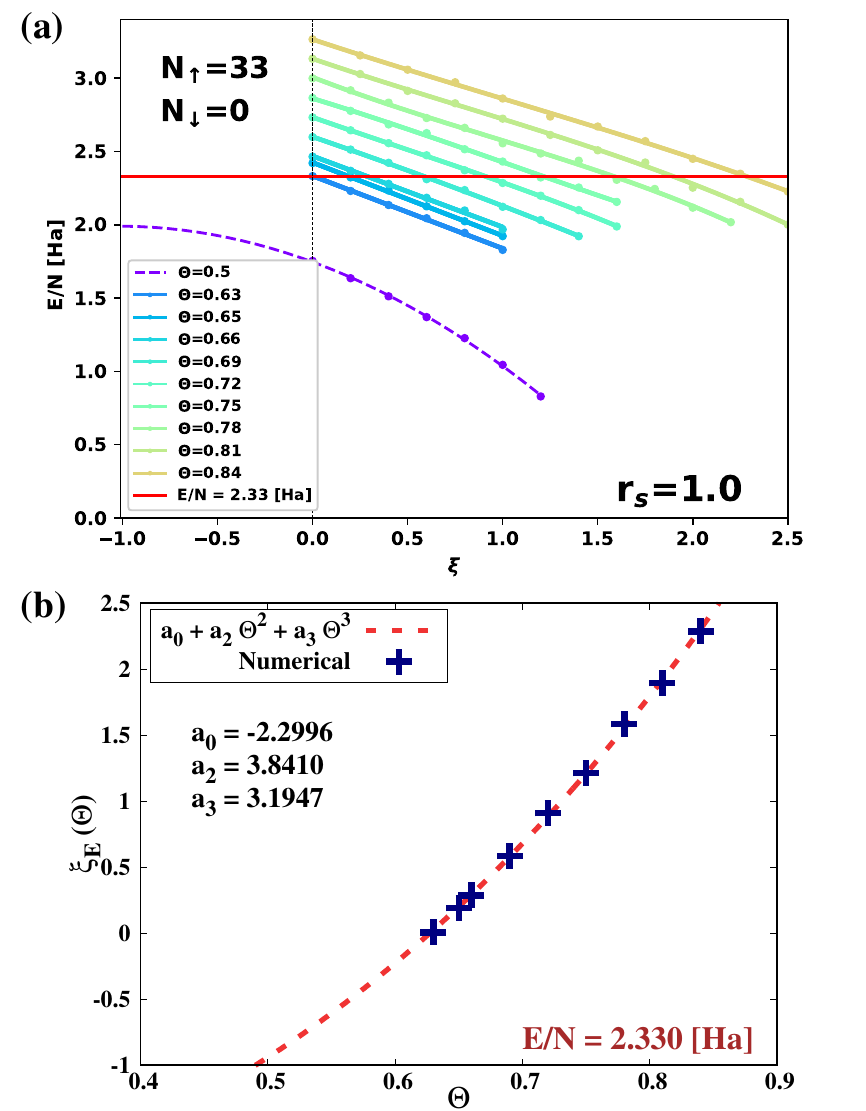}
\caption{(a) Energy per particle for $r_\text{s}=1.0$, $N=33$ and $\zeta=1$ as a function of the $\xi$-parameter for different $\Theta$. The red continuous line corresponds to the constant energy solution using CPIMC for $\Theta=0.5$. Purple dashed line is the fit to the energy values for $\Theta=0.5$. (b) Constant energy extrapolation of $\xi_\text{E}(\Theta)$ obtained by cutting the $E_\text{T}(\xi)/N$ functions in (a) along the red line.}\label{fig:rs_1}
\end{figure}

In Fig.~\ref{fig:rs_1} we report the result for $r_\text{s}=1.0$, $N=33$ and $\zeta=1$. In Fig.~\ref{fig:rs_1}(a) we show the behaviour of the energy-per-particle as a function of $\xi$. 

The constant-energy extrapolation shown in Fig.~\ref{fig:rs_1}(b) at the given $E/N=2.33$ Ha, which is again the benchmark result obtained using CPIMC~\cite{groth_2016}, recovers the true value for $\Theta$ with an uncertainty less than $2\%$. We notice again a smooth behaviour of $\xi_E(\Theta)$, which is fitted using Eq.~(\ref{eq:csi_exp_xiong}) up to the cubic term.
As in the previous subsection~(\ref{sec:rs_0.5}) instead, the constant-temperature extrapolation (purple dashed line in (a)) is not able to accurately predict the correct value of the energy-per-particle (red-line) for $\Theta=0.5$.

\subsubsection{UEG for $r_\text{s}=10.0$, $\Theta=0.5$, $N=33$ and $\zeta=1$}\label{sec:rs_10.0}
\begin{figure}[!hbt]
\centering
\includegraphics[width=0.5\textwidth]{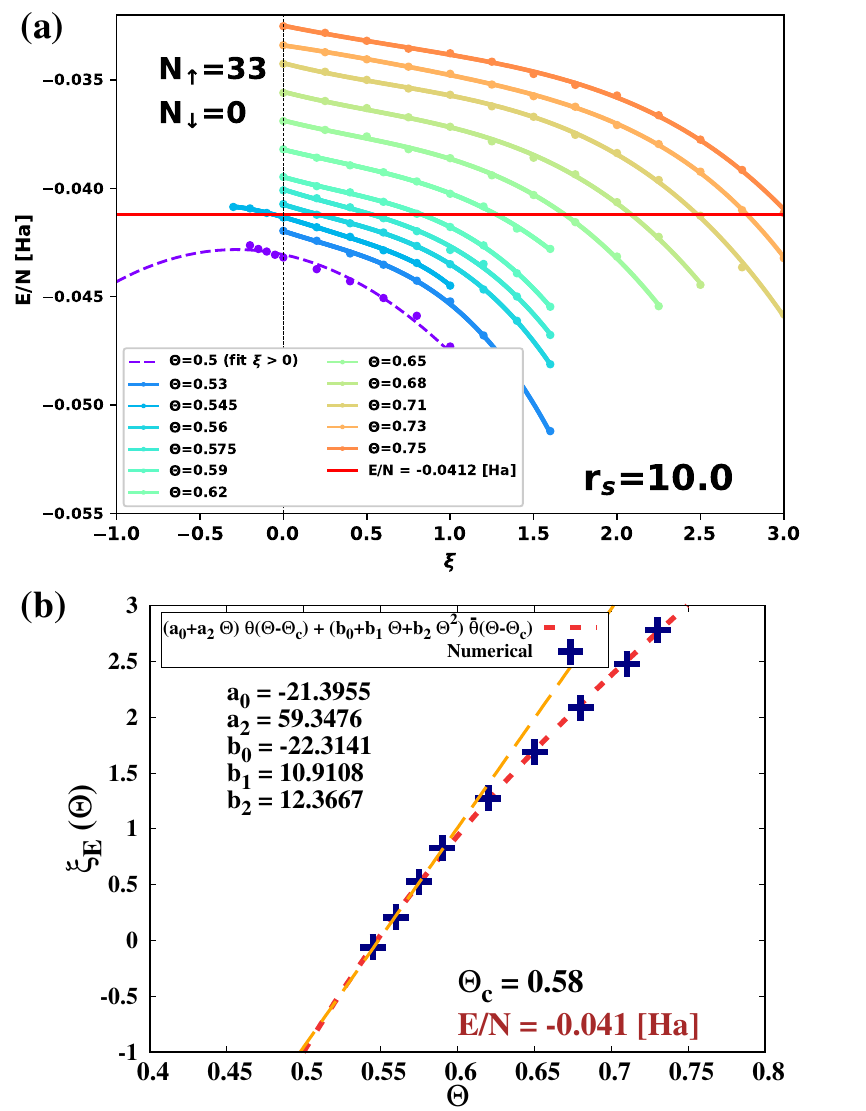}
\caption{(a) Energy per particle for $r_\text{s}=10.0$, $N=33$ and $\zeta=1$ as a function of the $\xi$-parameter for different $\Theta$. The red continuous line corresponds to the constant energy solution using CPIMC for $\Theta=0.5$. Purple dashed line is the fit to the energy values for $\Theta=0.5$. (b) Constant energy extrapolation of $\xi_\text{E}(\Theta)$ obtained by cutting the $E_\text{T}(\xi)/N$ functions in (a) along the red line. The orange dashed line indicates the linear fit applied in the regime where $\Theta < \Theta_c$. }\label{fig:rs_10}
\end{figure}

In Fig.~\ref{fig:rs_10} we report the result for $r_\text{s}=10.0$, $N=33$ and $\zeta=1$. Like in the previous two sections, in Fig.~\ref{fig:rs_10}(a) we show the behaviour of the energy-per-particle as a function of $\xi$.
In contrast to the case where $ r_s = 1.0 $, and as anticipated from the independent particle model analysis, the transition to BEC occurs at a smaller value of $ \xi$. This is evident from the change in the curvature of the energy plot. As noted in Ref.~\cite{morresi_2025} the observed smooth change in curvature is a finite-size effect. 

In Fig.~\ref{fig:rs_10}(b) we display the behaviour of $\xi_E(\Theta)$ for the given energy $E/N=-0.0412$ Ha, corresponding to the PB-PIMC result reported in Ref.~\cite{groth_2016}. As expected from the independent particle model for the plasmon, the trend is totally different with respect to the case $r_s=1.0$. Therefore, to extract the temperature corresponding to the fixed energy, we utilize Eq.~(\ref{eq:csi_exp_mine_linear}). This behavior closely resembles the shape observed in Ref.~\cite{morresi_2025} for liquid helium-3. Additionally, we have added a point at small negative $\xi$, which corresponds to the $\xi$ at $\Theta = 0.545$ in Fig.~\ref{fig:rs_10}(a), where we also present negative $\xi$ points for that temperature. This point has been added to validate the shape of the $\xi_E(\Theta)$ function and highlights that the small negative $\xi$ region can be considered when the sign problem is not severe. 
We will explain better the use of the small negative $\xi$ region in the following subsection.
We found that the temperature obtained through this extrapolation is approximately $\Theta = 0.5$, with an uncertainty of $1.5\%$.
We also stress that, as in the case of $r_s=1.0$, in Fig.~\ref{fig:rs_10}(a) the constant-temperature extrapolation at $\Theta=0.5$ (purple dashed line) for $\xi \geq 0$ does not reproduce the energy value given by the red horizontal line. 

\subsubsection{UEG for $r_\text{s}=80.0$, $N=33$ and $\zeta=1$}\label{sec:rs_80.0}
\begin{figure*}[!hbt]
\centering
\includegraphics[width=0.48\textwidth]{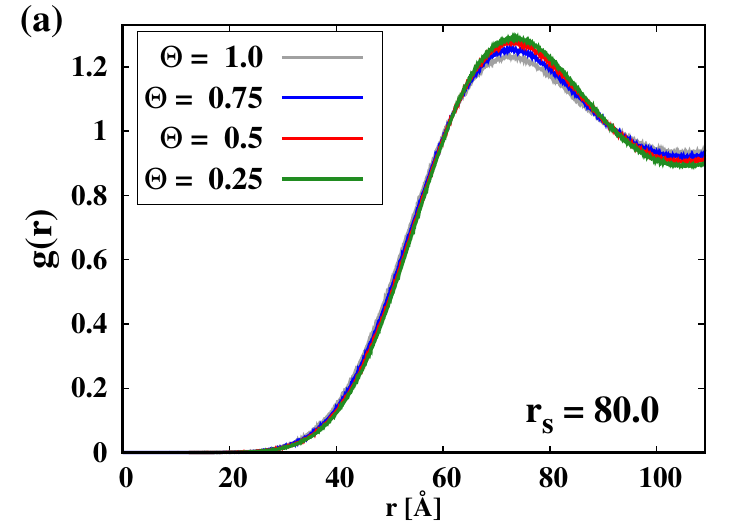}
\includegraphics[width=0.48\textwidth]{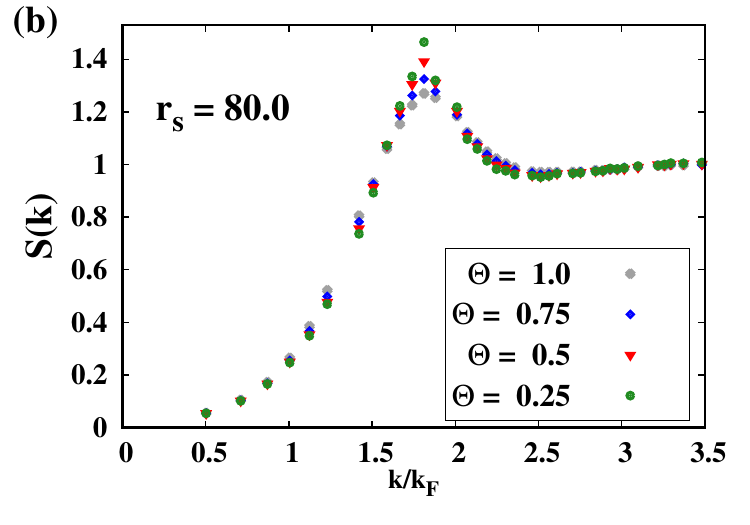}
\includegraphics[width=0.48\textwidth]{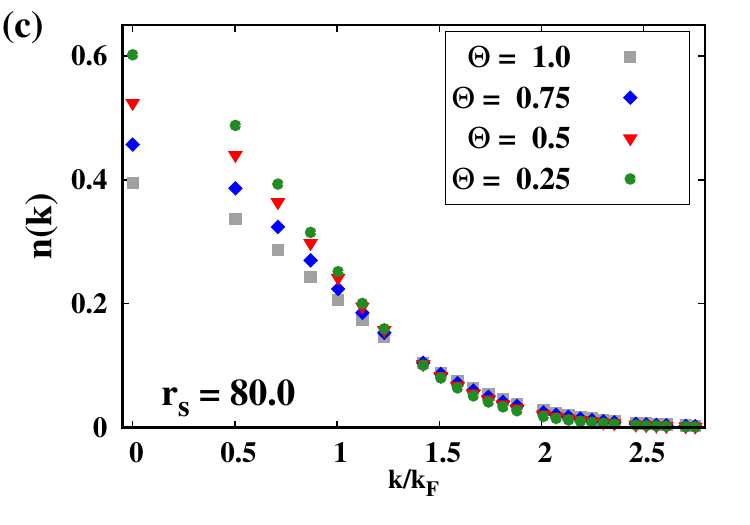}
\includegraphics[width=0.48\textwidth]{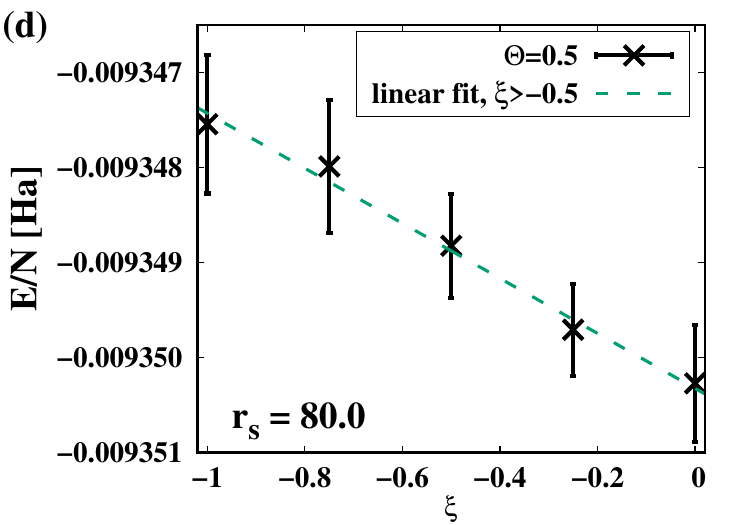}
\caption{(a) Pair correlation function for $r_\text{s}=80.0$, $N=33$, $\zeta=1$. The grey line corresponds to $\Theta=1.0$, blue line to $\Theta=0.75$, red line to $\Theta=0.5$ and green line to $\Theta=0.25$. (b) Static structure factors. Colors are the same as in (a). (c) Momentum distribution for the same four different temperatures as in (a) and (b). (d) Energy per particle for $\Theta=0.5$. The dashed line represents the linear fit taking into account only the three points for $\xi \geq -0.5$. The points at $\xi=-0.75$ and $\xi=-1$ are added to show that they fall into the line extrapolated from the fit at $\xi \geq 0$.}\label{fig:rs_80}
\end{figure*}

In the UEG, with fixed number of particles and temperature, the sign problem is particularly pronounced at high densities. However, it becomes more manageable as $r_s$ increases. In this subsection, we focus on $r_s = 80.0$, a region where, at least at zero temperature ($T = 0$), the polarized state is expected to be favored over the unpolarized state~\cite{ceperley_1980}. Using this example, we will demonstrate the effectiveness of the fictitious identical particles framework within the small negative $\xi$ sector. In particular, for $N=33$ polarized electrons at $r_s=80.0$ and $\Theta>0.25$ the sign problem is not severe~\cite{dornheim_2019}; therefore we can also simulate energy for $\xi=-1$ by direct PIMC. The purpose of this subsection is to show that a few points with $-0.5<\xi<0$ are enough to extrapolate the correct results at $\xi=-1$. 

Indeed, the key concept is that the BEC transition in UEG ---but also in general--- always occurs for $\xi > 0$, a behavior that can also observed in independent particle models. Consequently, all non-analytic behavior of the energy is also confined to the region where $\xi > 0$. In particular, when the BEC transition occurs near $\xi \sim 0^+$, as seen in this parameters regime, conducting simulations for $\xi > 0$ becomes challenging. This difficulty arises from the extensive permutation sampling required to accurately calculate the kinetic energy. In this scenario, exploring the region of small negative $\xi$ is often more favorable, as long as the sign problem is not too severe to prohibit direct PIMC simulations.

In Figs.~\ref{fig:rs_80}(a) and (b), based on direct PIMC simulations, we show the results obtained at different temperatures for $r_s=80.0$ for the pair correlation function $g(r)$ and the static structure factor $S(k)$ respectively. 
We simulate the static structure factor based on Eq. (2) in Ref. \cite{zhang2016concept}. 
It turns out that in the $[-1,0]$ range of $\xi$, the variation of $g(r)$ is weak and below $0.5\%$ in the region where $g(r)$ tends to $1$ (see Fig.~\ref{fig_si:gr_diff} in the Supplemental Material~\cite{supp}). Therefore, the structural properties of the UEG in this range of the density do not depend significantly on quantum statistics.
We notice that by decreasing the temperature, the main peaks of $g(r)$ and $S(k)$ become sharper. 
In particular, the behavior of $S(k)$ is characteristic of the strongly coupled electron liquid regime.

In Fig.~\ref{fig:rs_80}(c), we present the results for the momentum distribution $n(k)$. This function displays the occupation of momentum states which can be highly non-trivial for temperatures larger than zero~\cite{militzer_2002}. Using direct PIMC one can access the exact behaviour of $n(k)$~\cite{dornheim_nk}. In this work, the momentum distribution is obtained by Fourier transform of the one-body density matrix (see Eq.~(\ref{eq:obdm}) in the Supplemental Material~\cite{supp}). At variance with the pair correlation and structure factor, $n(k)$ depends on the $\xi$-parameter. However, this dependency is mild and a linear extrapolation gives of points for which $-0.5<\xi<0$ reproduces very well the limit of $\xi=-1$. This is shown in Fig.~\ref{fig_si:rs_comp}(a) where we reconstruct the whole momentum distribution for $\Theta=0.5$ and in Figs.~\ref{fig_si:rs_comp}(b) and (c) in the Supplemental Material \cite{supp}, where for $\Theta=0.5$ we zoomed over the smallest q-points sampled by our simulation.

Finally, in Fig.~\ref{fig:rs_80}(d) we show the behaviour of the energy-per-particle at constant temperature, that is $\Theta=0.5$. We observe a variation of the energy-per-particle of the order of $\sim0.1$ meV from the Boltzmannon ($\xi=0$) to the Fermionic case ($\xi=-1$), meaning that quantum statistical effects are marginal for the energy. However, the behaviour in the negative-$\xi$ sector is quite regular as expected, and the fermionic energy can be inferred by using only the points for $-0.5<\xi<0$. This is shown by the dashed line which is a linear fit for $-0.5<\xi<0$.

\section{Discussion and Conclusion}
\label{DC}
In this paper we have applied the framework of fictitious identical particles~\cite{xiong_2022,xiong_2023,dornheim_1_2023,dornheim_1_2024,dornheim2024ab, dornheim2024ab1, dornheim2024unraveling,morresi_2025,xiong2024quadratic,xiong2024gpu,yang2025density,dornheim2025fermionic} to the UEG at different electronic densities, including the regime where degeneracy effects play a major role. We have shown that the constant-energy extrapolation~\cite{xiong_2023,morresi_2025} works better than the constant-temperature extrapolation advocated in Ref.~\cite{xiong_2022}. While both methods are of equivalent quality in the warm dense matter (WDM) regime ($\Theta \gtrsim  1$)~\cite{dornheim_1_2023,dornheim_1_2024,dornheim2024ab, dornheim2024ab1, dornheim2024unraveling}, the latter has been previously employed in literature under conditions where quantum degeneracy effects are weak. However, as temperature decreases, the constant-temperature extrapolation becomes ineffective, as demonstrated in sections 
\ref{sec:rs_0.5}, \ref{sec:rs_1.0}, and \ref{sec:rs_10.0}. 

The constant-energy extrapolation allows a more general description because the $\xi_E(\Theta)$ functions can be built in such a way to circumvent the region where the energy drops very rapidly to the ground state level.

With the help of an independent particle model, we have also demonstrated the effect of weak ($r_s \leq 1$) and strong ($r_s \geq 10$) correlations of the UEG in the context of the fictitious identical particles. Weak correlations correspond to an analytical behaviour of the $\xi_E(\Theta)$ functions that can be fitted with simple polynomials, while strong correlations make those functions non-analytical due to the sharper BEC transition.

In the framework of fictitious identical particles, we also put forward the idea of using the small-negative $\xi$ region where the sign problem turns out to be not severe. We find this approach very important for the present and future works, because the BEC transition is always happening for $\xi>0$ and therefore all the non-analyticity is confined in the range $\xi>0$. For $\xi\leq0$ a few points can be very useful to extrapolate more accurately to the fermionic limit of the observables of interest, as we have shown in subsection \ref{sec:rs_80.0}. It is worth pointing out that a recent work by Dornheim \textit{et al.} \cite{dornheim2025fermionic} used a similar approach in the framework of the 
$\eta$-ensemble PIMC method, obtaining highly accurate values of the fermionic free energy in the warm dense UEG.

We believe that the fictitious identical particle method, being fully \textit{ab initio} and exact in the region where direct calculations are performed, can be further extended to address several many-body challenges in quantum chemistry and, in the long term, contribute to tackling high-temperature superconducting systems.

\bibliography{biblio}

\newpage
\begin{CJK*}{}{} 
\author{Tommaso Morresi$^{1,2\ast}$, Giovanni Garberoglio$^{2}$, Hongwei Xiong$^{1}$, Yunuo Xiong$^{1\ast}$\\~\\}
\affiliation{$^{1}$ Center for Fundamental Physics and School of Mathematics and Physics, Hubei Polytechnic University, Huangshi 435003,
People's Republic of China}
\affiliation{$^{2}$ European Centre for Theoretical Studies in Nuclear Physics and Related Areas (ECT*), Fondazione Bruno Kessler, Italy}
\title{\LARGE{SUPPLEMENTAL MATERIAL}\\Study of the uniform electron gas through parametrized partition functions}

\maketitle
\end{CJK*}
\setcounter{section}{0}
\setcounter{figure}{0}
\setcounter{equation}{0}
\renewcommand{\theequation}{S.\arabic{equation}}
\renewcommand{\thefigure}{S.\arabic{figure}}
\renewcommand{\thetable}{S.\arabic{table}}
\renewcommand{\thesection}{S.\arabic{section}}
\renewcommand*{\citenumfont}[1]{S#1}
\renewcommand*{\bibnumfmt}[1]{[S#1]}
\onecolumngrid

\section{Estimators}\label{sec:estim}
In the primitive approximation, the virial estimator for the energy is computed as [S1, S2]
\begin{equation}\label{eq:ev}
\begin{split}
  E_V =& \Big\langle \frac{3N}{2\beta} +\frac{\left( \mathbf{R}^{(P-1)}-\mathbf{R}^{(P)}\right)\cdot \left( \mathbf{R}^{(P)}-\mathbf{R}^{(0)}\right)}{2\tau^2P} 
  + \frac{1}{P} \sum_{j=0}^{P-1} V \left(\mathbf{R}^{(j)} \right) \\
  &+ \frac{1}{2P} \sum_{j=0}^{P-1} \left( \mathbf{R}^{(j)} -\mathbf{R}^{(0)} \right) \cdot \frac{\partial V \left(\mathbf{R}^{(j)} \right)}{\partial \mathbf{R}^{(j)}} \Big\rangle,    
\end{split}
\end{equation}
where the brackets imply the average over the configurations generated through PIMC and $P$ is the number of beads. 

The static structure factor is not computed using the Fourier transform of the $g(r)$ but using the direct formula, which reads as [S3]: 

\begin{equation}
    S(\mathbf{k}) = \frac{2}{N} \sum_{i=1}^N\sum^N_{k>i} \left[\text{cos}\left(\mathbf{k}\cdot \mathbf{x}^{(0)}_i\right) \text{cos}\left(\mathbf{k}\cdot \mathbf{x}^{(0)}_k\right) + \text{sin}\left(\mathbf{k}\cdot \mathbf{x}^{(0)}_i\right) \text{sin}\left(\mathbf{k}\cdot \mathbf{x}^{(0)}_k\right) \right],
\end{equation}
where $\mathbf{x}_i^{(0)}$ labels the position of the \textit{i}-th atom in the first time slice inside the unit cell. 

The momentum distribution is instead computed from the Fourier transform of the one-body density matrix (OBDM) $\rho_1(\mathbf{r},\mathbf{r}')$.
In the context of PIMC + worm algorithm, $\rho_1(\mathbf{q})$ with $\mathbf{q}=\mathbf{r}-\mathbf{r}'$ can be estimated as
\begin{equation}\label{eq:obdm}
    \rho_1 (\mathbf{q}) = \frac{\Omega}{N Z} \left\langle \delta \left( \mathbf{q} - \left( \mathbf{r}^{(0)}_{i_T} - \mathbf{r}^{(P)}_{i_H} \right) \right)\right\rangle,
\end{equation}
where $\mathbf{r}^{(0)}_{i_T}$ is the position of the tail of the worm, $\mathbf{r}^{(P)}_{i_H}$ is the position of the head of the worm and the brackets here refer to the average over the configurations presenting the open polymer (G-sector) [S4]. 
In particular, the OBDM should meet the normalization condition $\rho_1(0)=1$. In practice, in our implementation the OBDM is spherically symmetric and it is estimated as:
\begin{equation}
    \rho_1(q) = \frac{1}{nCN_Z\Omega_q}\langle q - \left| \mathbf{r}^{(0)}_{i_T} - \mathbf{r}^{(P)}_{i_H} \right| \rangle,
\end{equation}
where $n$ is the density of electrons, $C$ is the open-close parameter appearing in the acceptance/rejection of the open and close moves, $N_Z$ is a counter for the number of times the system is in the Z-sector (the one with all closed polymers) [S2].

\section{A three-dimensional perspective on extrapolation}
\begin{figure}[!hbt]
\centering
\includegraphics[width=0.48\textwidth]{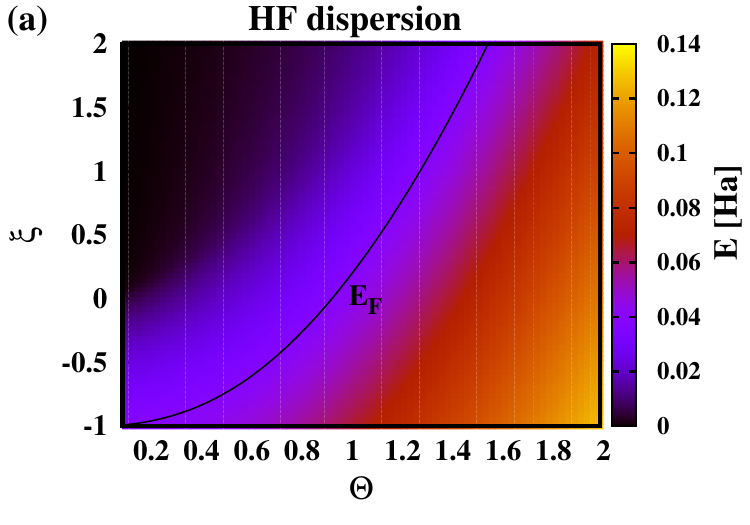}
\includegraphics[width=0.48\textwidth]{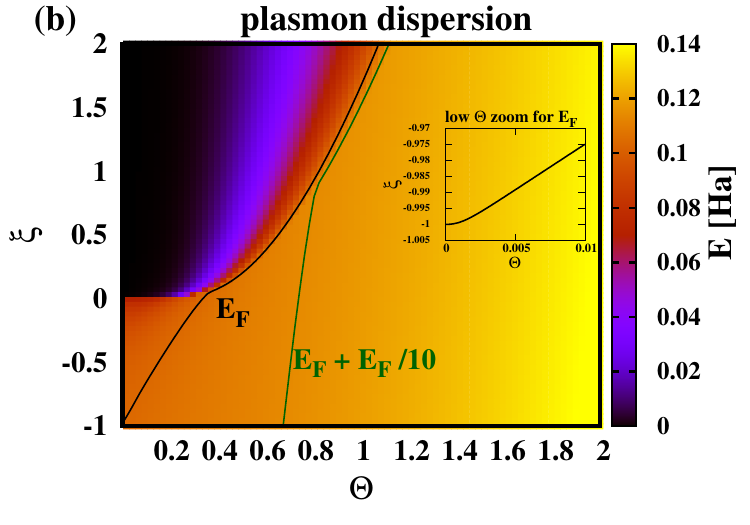}
\caption{Energy as a function of $\xi$ and $\Theta$ for (a) the HF dispersion in Eq.~(\ref{eq:disp_hf}) and (b) the plasmon dispersion in Eq.~(\ref{eq:disp_helium-like}) of the main text. In both cases, $r_s=10.0$. The continuous black lines in both plots correspond to the constant energy at the Fermi level, while the dark-green line in (b) correspond to a higher energy. The inset in (b) is a zoom over $\Theta=0$.}\label{fig_si:3d}
\end{figure}

In this section, we aim to elucidate the differences between constant-temperature and constant-energy extrapolations, as discussed in Sec. \ref{CD} of the main text. To illustrate these differences, we utilize the two independent particle models represented by Eqs.~(\ref{eq:disp_helium-like}) and (\ref{eq:disp_hf}) in the main text, where we fixed $r_s=10.0$. In Fig. \ref{fig_si:3d}, we present the energy as a function of $\xi$ and $\Theta$ for both dispersion relations. Specifically, panel (a) focuses on the HF dispersion (Eq. (\ref{eq:disp_hf})), while panel (b) examines the plasmon dispersion (Eq. (\ref{eq:disp_helium-like})). Notably, the energy behavior in panel (a) is significantly smoother compared to that in panel (b), where the inflection point of the energy (see Fig.~\ref{fig:ipm_pl}(c) of the manuscript) is evident from the sharp colour change at low temperature.

However, in both cases for low temperatures, performing a constant-temperature extrapolation by fitting the energies at $\xi \geq 0$ to extract the fermionic energy becomes challenging. This difficulty arises due to the drop in energy associated with the BEC formation, signaled by the black colour in both panels. As the temperature decreases, the critical value $\xi_c$ at which the BEC transition occurs also decreases. As a result, the energy function undergoes an abrupt change of curvature, complicating the extrapolation process due to the inherent non-analyticity associated with BEC.

Instead, the constant-energy extrapolation circumvents the "black hole" region where the energy drops to the ground state. In Fig.~\ref{fig_si:3d}(a) the behaviour of the $\xi_E(\Theta)$ is quite regular, as can be seen from the black line, representing the Fermi energy. Therefore, in this case, the energies can be easily extrapolated using the constant-energy curves.
In Fig.~\ref{fig_si:3d}(b), the functions $\xi_E(\Theta)$ also avoid the "black hole" region; however, their behavior is more complex due to the sharper transition.

We observe that the apparent violation of the condition $\frac{\partial \xi_E}{\partial \Theta}|_{\Theta=0}=0$ (see Ref. [S5]) is merely a graphical issue. In fact, the inset in Fig.~\ref{fig_si:3d}(b) provides a zoomed-in view of the region where $\Theta \to 0$, clearly showing that the derivative of $\xi_E(\Theta)$ is correctly recovered as $\Theta$ approaches 0.

\section{Extrapolation convergence of the UEG for $r_\text{s}=80.0$, $N=33$ and $\zeta=1$}\label{sec:rs_80_SI}
\begin{figure}[!hbt]
\centering
\includegraphics[width=0.9\textwidth]{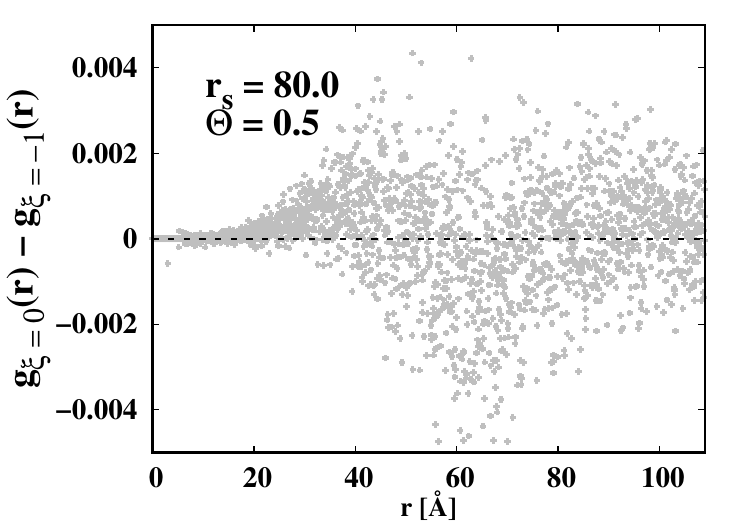}
\caption{Difference between the pair correlation function for $\xi=0$ and $\xi=-1$ in the case of $r_s=80.0$ and $\Theta=0.5$.}\label{fig_si:gr_diff}
\end{figure}

In Fig.~\ref{fig_si:gr_diff}, we show the difference between the pair distribution function at $\xi = 0$ and that at $\xi = -1$ for $\Theta = 0.5$. Our findings indicate that in the region where $ g(r) $ is non-zero (specifically for $ r \gtrsim 30 $), the differences are symmetrically distributed around zero. Furthermore, the absolute values of these differences do not exceed  $5 \cdot 10^{-3} $. This suggests that the structural properties of the system are largely unaffected by quantum statistics, highlighting the stability of the system's behavior despite variations in the parameter $ \xi $.

\begin{figure}[!hbt]
\centering
\includegraphics[width=0.99\textwidth]{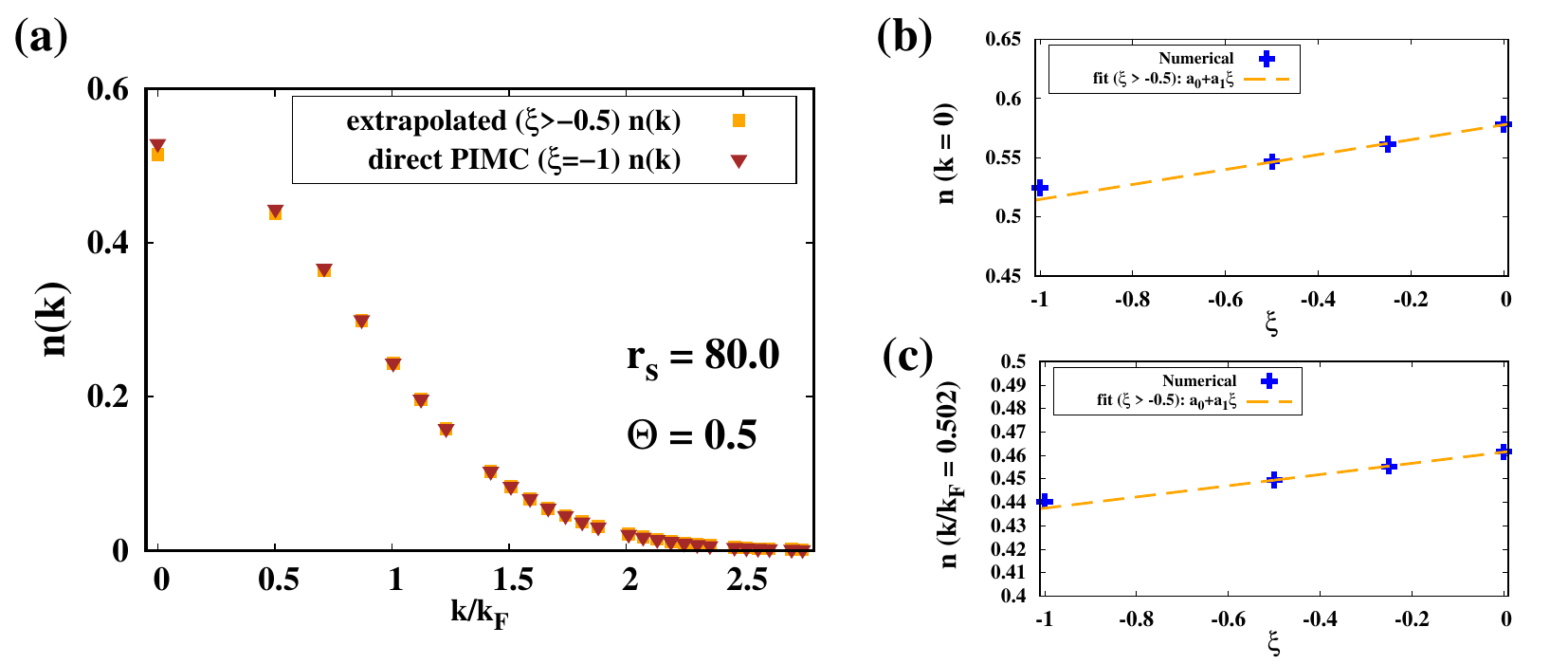}
\caption{(a) Momentum distribution extrapolated using data with $\xi \geq-0.5$ (orange squares) versus direct PIMC calculation for $\xi=-1$ (brown triangles). (b) Linear extrapolation of the point at $k=0$ using points at $\xi \geq 0$. (c) Linear extrapolation of the point at $k/k_\text{F}=0.502$ using points at $\xi \geq 0$. }\label{fig_si:rs_comp}
\end{figure}

In Fig.~\ref{fig_si:rs_comp}, we present the momentum distribution for $\Theta = 0.5$. Panel (a) of this figure compares the direct PIMC calculation (represented by brown triangles) with the extrapolated momentum distribution $n(k)$ (orange squares), using data points where $\xi > -0.5$. Notably, a small discrepancy is observed at the point where $k = 0$. This discrepancy is further highlighted in Fig.~\ref{fig_si:rs_comp}(b), where we provide a zoomed-in view along with its linear extrapolation. For all other points where $k > 0$, the results are nearly indistinguishable, indicating a strong agreement between the direct calculation and the extrapolated values.

\clearpage
\section*{References}
\begin{enumerate}
\itemsep-0.2em 
\small
    \item[ \text{[S1]} ] D. M. Ceperley, Path integrals in the theory of condensed helium, Rev. Mod. Phys. 67, 279 (1995).
    \item[ \text{[S2]} ] G. Spada, S. Giorgini, and S. Pilati, Path-Integral Monte Carlo Worm Algorithm for Bose Systems with Periodic Boundary Conditions, Condensed Matter 7 (2022).
    \item[ \text{[S3]} ] K. Zhang, On the concept of static structure factor, arXiv:1606.03610 (2016).
    \item[ \text{[S4]} ] T. Morresi and G. Garberoglio, Revisiting the properties of superfluid and normal liquid $^4$He using ab initio potentials, Journal of Low Temperature Physics 219, 103 (2025).
    \item[ \text{[S5]} ] Y. Xiong and H. Xiong, Thermodynamics of fermions at any temperature based on parametrized partition function, Phys.
Rev. E 107, 055308 (2023).
\end{enumerate}

\end{document}